\documentclass[twocolumn,showpacs,preprintnumbers,amssymb,prl,nofootinbib]{revtex4}
\usepackage[utf8]{inputenc} 
\usepackage[english]{babel}
\usepackage[T1]{fontenc}
\usepackage{graphicx,amssymb,amsmath,wasysym}
\usepackage[bookmarks=false]{hyperref}
\usepackage{graphicx}
\usepackage{dcolumn}
\usepackage{bm}

\def\gtap{\mathrel{ \rlap{\raise 0.511ex \hbox{$>$}}{\lower 0.511ex
   \hbox{$\sim$}}}} 
\def\ltap{\mathrel{ \rlap{\raise 0.511ex
    \hbox{$<$}}{\lower 0.511ex \hbox{$\sim$}}}} 
\newcommand{\bea}{\begin{eqnarray}} 
\newcommand{\eea}{\end{eqnarray}}
\def\beq{\begin{equation}}
\def\enq{\end{equation}}
\def\ba{\begin{eqnarray}}
\def\ea{\end{eqnarray}}
\def\<{<\!\!}
\def\>{\!\!>}
\def\<{\langle}
\def\>{\rangle}

\begin{document}

\input{epsf}

\preprint{CFTP/11-018}
\preprint{IPPP/11/63}
\preprint{DCPT/11/126}
                 
\title{Determining the dark matter mass with DeepCore}

\author{Chitta R. Das$^1$, Olga Mena$^2$, Sergio Palomares-Ruiz$^1$
  and Silvia Pascoli$^3$} 

\affiliation{$^1$Centro de Física Teórica de Partículas,
Instituto Superior Técnico, Universidade Técnica de Lisboa, Avenida
Rovisco Pais 1, 1049-001 Lisboa, Portugal}
\affiliation{$^2$Instituto de Física Corpuscular (IFIC), 
CSIC-Universitat de València, Apartado de Correos 22085, E-46071
Valencia, Spain}
\affiliation{$^3$IPPP, Department of Physics, Durham University,
  Durham DH1 3LE, United Kingdom} 

\begin{abstract}
Cosmological and astrophysical observations provide increasing
evidence of the existence of dark matter in our Universe.  Dark matter
particles with a mass above a few GeV can be captured by the Sun,
accumulate in the core, annihilate, and produce high energy neutrinos
either directly or by subsequent decays of Standard Model particles.
We investigate the prospects for indirect dark matter detection in the
IceCube/DeepCore neutrino telescope and its capabilities to determine
the dark matter mass. 
\end{abstract}

\maketitle

{\bf Introduction.---} 
Establishing the particle identity of the dark matter (DM) of the
Universe is one of the fundamental questions which needs to be
addressed in modern physics~\cite{review}.  One of the currently
favored candidates is a weakly interacting massive particle (WIMP), a
stable neutral particle with a mass in the GeV--TeV range, which is
predicted in many extensions of the Standard Model of particle physics
(SM).

The DM particle properties might be tested in direct and indirect
searches and in collider experiments.  Direct searches look for the
recoil of nuclei due to WIMPs passing through the detector and are
sensitive to the spin-dependent WIMP-proton (-neutron) cross section,
$\sigma_{\mathrm{SD}}^\mathrm{p (n)}$, and the spin-independent one,
$\sigma_{\mathrm{SI}}$.  Indirect searches focus on the products of DM
annihilations (or decays), such as gamma-rays, positrons, anti-protons
and neutrinos, in regions where the DM density is expected to be high,
such as the center of the Milky Way, dwarf galaxies, clusters of
galaxies, etc.  Colliders could produce DM particles and make possible
the study of their properties.  In addition to each individual search,
their combination is crucial in constraining DM
properties~\cite{DMmassgammas, DMmassdirect, DMmassneutrinos,
  Cirelli:2005gh, DMmasscollider}.  In particular, the determination
of the DM mass is of great theoretical importance in order to
establish the DM particle identity but it is a challenging task. 

\begin{figure}[t]
\begin{center}
\includegraphics[width=0.95\linewidth]{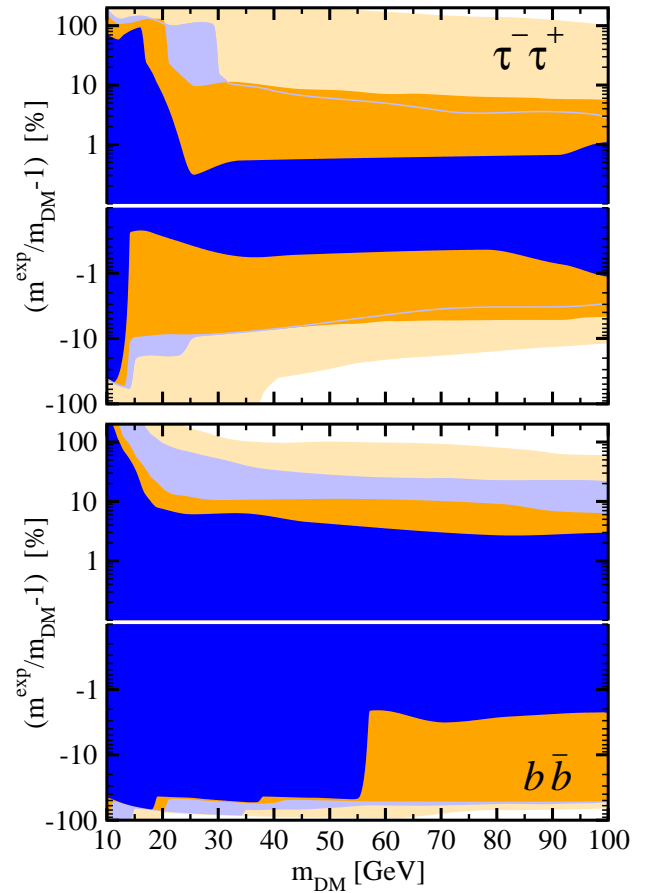}
\caption{\sl \footnotesize \textbf{\textit{Relative error in the DM
      mass after 10 years at DeepCore at 90\% CL}}, with (light
  colors) and without (dark colors) systematic errors. Top panel:
  Annihilations into $\tau^-\tau^+$ with
  $\sigma_{\mathrm{SD}}^\mathrm{p} = 10^{-3}$~pb (blue), $10^{-4}$~pb
  (orange). Bottom panel: Annihilations into $b\bar{b}$ with
  $\sigma_{\mathrm{SD}}^\mathrm{p} = 10^{-2}$~pb (blue), $4 \times
  10^{-3}$~pb (orange).}
\label{fig:fig1}
\vspace{-8mm}
\end{center}
\end{figure}

One of the astrophysical objects of particular interest for indirect
searches is the Sun as a high DM density could be present: DM
particles, traversing it, can get scattered to velocities lower than
the escape velocity and be captured.  The annihilations of these DM
particles would give rise to a neutrino flux, which is produced either
directly or indirectly via the decay of other products of the
annihilations.  Depending on the annihilation channel, the neutrino
energy is different: neutrinos from particles which decay very fast
have typically high energies and longer lived particles, such as muons
and light quarks, significantly loose energy before decaying and
produce softer neutrinos~\cite{Cirelli:2005gh}.  Therefore, the
neutrino spectrum depends on several DM properties, such as its mass,
the DM-nucleon cross section and the branching ratios for the various
possible annihilation channels.

Indirect searches, detecting these neutrinos and reconstructing their
spectrum, are sensitive to the neutrino mass~\cite{Cirelli:2005gh}.
It has been shown that future large neutrino detectors with good
energy resolution, such as magnetized iron detectors, could measure
the {\it neutrino} spectrum~\cite{Mena:2007ty, Agarwalla:2011yy}.
However, so far, no simulation of this effect in a neutrino telescope
has been performed. These very large detectors cannot fully
reconstruct the neutrino energy but measure the {\em neutrino-induced}
muon spectrum, which is strongly correlated with the neutrino one.
Here, we show for the first time (see Fig.~\ref{fig:fig1}) that, by
studying the neutrino-induced muon energy spectrum, the DeepCore Array~\cite{Collaboration:2011ym}, a huge compact \v{C}erenkov detector located at the bottom center of the IceCube Neutrino Telescope,  could reach an excellent precision in determining the DM mass.  This detector extends the IceCube neutrino detection capabilities to neutrino energies as low as 10~GeV allowing to study
low mass WIMPs with a detector with a huge effective volume,
$\mathcal{O}$(10)~Mton.

{\bf Neutrinos from WIMPs annihilations in the Sun.---} 
When a DM particle with a mass above a few GeV passes through the Sun,
it might interact elastically with the nuclei and get scattered to a
velocity smaller than the escape velocity, remaining gravitationally
trapped.  Then it undergoes additional scatterings, settling in
the Sun core, giving rise to an isothermal distribution.  For
sufficiently high capture rate and annihilation cross section,
equilibrium is reached and the annihilation rate
$\Gamma_{\mathrm{ann}}$ is related to the capture rate $C_\odot$
as~\cite{Gould92, Mena:2007ty}
\begin{eqnarray}
\Gamma_{\mathrm{ann}} \simeq \frac{C_\odot}{2} & \simeq &
\frac{9}{2} \times 10^{24} \; {\rm s}^{-1} \, 
\times \left(\frac{\sigma}{10^{-2} \, {\rm pb}}\right) 
    \; \left(\frac{50 \, {\rm GeV}}{m_{\rm DM}}\right)^2
 \nonumber
\\  
& & \left(\frac{\rho_{\rm local}}{0.3 \, {\rm GeV/cm}^{3}}\right) \,
\left(\frac{270 \, {\rm km/s}}{\bar{v}_{\rm local}}\right)^3 ~,
\label{eq:capture}
\end{eqnarray}
where $\rho_{\rm local}$ is the local DM density, $\bar{v}_{\rm
local}$ is the DM velocity dispersion in the halo and
$\sigma$ is the DM-nucleon cross section.  The spin-independent
scattering cross section is very strongly constrained by direct
searches~\cite{limitsdirect} and hence, at the energies of interest,
only signals due to a large spin-dependent cross section could be
tested with neutrino telescopes.  In many extensions of the SM the
spin-dependent cross section can dominate, even by several orders of
magnitude~\cite{SDvsSD}.  The current most stringent bounds on the
WIMP-proton elastic scattering cross section, $\sigma_{\mathrm
  {SD}}^\mathrm{p}$, are provided by the SIMPLE
experiment~\cite{Felizardo:2011uw}, whose Stage 1 and 2 combined 
results (only Stage 1 revised results) yield a lower limit with a 
minimum of $\sigma_{\mathrm{SD}}^\mathrm{p} < 4.2 \, (8.3) \times
10^{-3}$~pb at $m_{\rm DM} = 35$~GeV. This limit is already competitive
with the indirect ones obtained from the non-observation of neutrinos
from DM annihilations in the Sun at the Super-Kamiokande
detector~\cite{SKlimit, Kappl:2011kz}.

The DM accumulated in the Sun can annihilate into SM particles.
Nevertheless, due to the absorption in the solar matter, among all the
SM products of annihilations, only neutrinos can escape.  Neutrinos
could be produced either directly or after the hadronization,
fragmentation and decay of the SM particles in the final states.  A
broad spectrum of neutrinos arises and depends on the DM mass and on
the branching ratios into the various channels: 
\begin{equation}
\frac{dN_\nu}{d\Omega\ dt\ dE_\nu}=\frac{\Gamma_{\mathrm{ann}}}{4\pi
  R^2}\sum_i {\rm BR}_i \frac{dN_{i}}{dE_\nu}~, 
\label{eq:fluxes}
\end{equation}
where the sum includes the possible annihilation channels with
spectrum $dN_i/dE_\nu$ and branching ratio ${\rm BR}_i$, and $R$ is
the Sun-Earth distance.  The annihilation channels can be
distinguished into {\em hard channels} producing highly energetic
neutrinos, typically due to fast-decaying SM particles and {\em soft
  channels}, which are due to SM particles which interact significantly
with the high density background in the Sun loosing significant
amounts of energy before decaying~\cite{Cirelli:2005gh,
  Blennow:2007tw, Barger:2007xf}.  As our benchmark channels, we 
consider annihilations into $\tau$ pairs for the hard case
and into $b$ quarks for the soft one.

Once produced in the core of the Sun after DM annihilations, neutrinos
propagate undergoing neutrino oscillations, absorption due to neutral
and charged current interactions, loss of energy due to neutral
current and regeneration, when $\tau$ leptons produced in the
interactions decay into secondary lower energy
neutrinos~\cite{Cirelli:2005gh, Barger:2007xf, Blennow:2007tw}.  In
order to simulate the WIMP signal at the detector, including all the 
above effects, we use the publicly available code
\texttt{WimpSim}~\cite{Blennow:2007tw}.

{\bf Neutrino detection in DeepCore.---} 
The main idea of the present letter is to study the neutrino-induced
muon energy spectrum in order to gain information on the DM
properties, and in particular about the DM mass.  Using neutrinos from
DM annihilations in the Sun to constrain DM properties has been
considered in different contexts~\cite{DMmassneutrinos,
  Cirelli:2005gh, Mena:2007ty, Agarwalla:2011yy}.  Here, we focus for
the first time on the capabilities of the DeepCore Array for low mass
WIMPs.

DeepCore~\cite{Collaboration:2011ym} is located at the bottom center
of the IceCube detector at a depth between 2100~m and 2450~m, avoiding a
horizontal layer of poor optical properties due to a high content of
dust.  The detector has a higher instrumentation density with 6
additional strings instrumented with phototubes with higher quantum
efficiency with respect to IceCube. The same phototubes are
used for the IceCube strings in the same volume.  The advantages of
DeepCore are multiple.  The ice at this depth is on average twice as
clear as the one above allowing to detect a larger number of
unscattered photons and to achieve a better pattern recognition and low
energy neutrino reconstruction.  The higher vertical density of
photosensors, 7 m instead of 17 m for IceCube, and the higher quantum
efficiency lead to a significant gain in sensitivity, up to a factor
of 6, especially for low energy neutrinos.  Finally, the remaining 
volume of the IceCube detector together with a horizontal region with
additional instrumentation at a depth of 1750~m -- 1850~m can be used
as an active veto for downgoing atmospheric muons, significantly
reducing this dominant background. 

Due to the lack of angular resolution for cascade events, in this work
we only consider muon-like events (upgoing and downgoing), with an
effective volume for the 86-string configuration (IC86) at trigger
(SMT3) and online filter level $V_{\mathrm{eff}} \simeq $~8~Mton at
$E_\nu \simeq 10-12$~GeV and $V_{\mathrm{eff}} \simeq $~45~Mton at
$E_\nu \simeq 100-200$~GeV~\cite{talk}.  It is important to note that
this estimate of the effective volume does not include analysis or
reconstruction efficiencies. 
 The IceCube Collaboration aims to
maintain a signal efficiency of well over 50\% for contained and
partially contained events~\cite{Collaboration:2011ym}, which we
approximately account for by scaling down the simulated events by a
factor of~2.  For muon-like events, the angular resolution of the
detector is expected to be much better than the average angle between
the incoming neutrino and the produced muon.  Hence, the Sun is
basically a point source for this detector at these energies and we
consider the atmospheric neutrino background integrated over a
half-cone aperture given by $\theta_{\rm rms} =
\sqrt{\frac{1\ \textrm{GeV}}{E_\nu}}$.  As for the energy resolution,
it has not been estimated yet, but it will rely on track length rather 
than track brightness.  Assuming the track estimation to be
good to 50 meters, we consider bins with a 10~GeV width in the muon
energy. We assume 10 years of data taking. 

In these searches, the main source of background is due to atmospheric
neutrinos.  In the energy region of interest, the absolute atmospheric
neutrino fluxes are known within $\sim$10\%-20\%, the major
contributors coming from hadron production and the primary cosmic ray
fluxes~\cite{Barr:2006it}.  We note, though, that this uncertainty
could be substantially reduced~\cite{SKlimit}.  However, other sources
of systematic errors, such as the astrophysical uncertainties in the
calculation of the capture rate, could also affect the
results~\cite{Serpico:2010ae}.  All in all, we add an overall 15\%
systematic error in our computations as a conservative assumption.

\begin{figure}[t]
\begin{center}
\includegraphics[width=\linewidth]{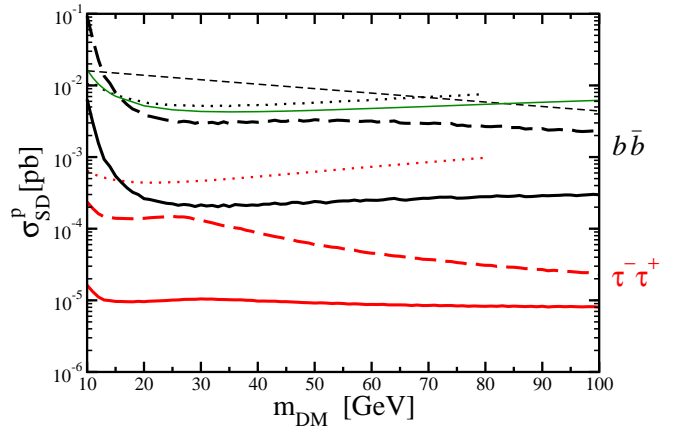}
\caption{\sl \footnotesize \textbf{\textit{Limits on the WIMP-proton
      spin-dependent cross section at 90\% CL.}} Black (upper) lines
  refer to annihilations into $b\bar{b}$ and red (lower) lines to
  annihilations into $\tau^-\tau^+$. Thick lines represent the limits
  expected with DeepCore after 10 years, including (dashed lines) and
  not including (solid lines) systematic errors.  The limit for the
  $b\bar{b}$ channel using stopping and through-going muons in
  Super-Kamiokande~\cite{SKlimit} is shown with the thin black dashed
  line.  Limits using fully-contained muon-like and upward stopping
  muons in Super-Kamiokande~\cite{Kappl:2011kz} are shown with the
  dotted lines.  The limits from the combination of the revised Stage
  1 and Stage 2 of the SIMPLE experiment are depicted by the green solid
  line~\cite{Felizardo:2011uw}. 
} 
\label{fig:fig2}
\vspace{-8mm}
\end{center}
\end{figure}

{\bf Determination of the DM mass in DeepCore.---} 
In Fig.~\ref{fig:fig2}, we show the sensitivity to DM annihilation in
the Sun due to elastic spin-dependent interactions off protons at 90\%
confidence level (CL).  We show the results for the $\tau^-\tau^+$
(hard) and $b\bar{b}$ (soft) annihilation channels.  For comparison,
we also show the recent results from the direct searches by the SIMPLE
experiment~\cite{Felizardo:2011uw} and those using Super-Kamiokande
data~\cite{SKlimit, Kappl:2011kz}.  It is important to note that the
results of these latter analyses do not include systematic errors of
the kind mentioned above, so should be compared with our solid lines
in Fig.~\ref{fig:fig2}.

The results of the present letter rely on the capability of DeepCore
to reconstruct the (muon) energy spectrum: distinguishing between hard
and soft channels allows to get information on the initial
annihilation channels, mass and WIMP-proton cross section.  If only
the total number of events is measured, a strong degeneracy is present
among these parameters~\cite{Mena:2007ty}.  In particular, here we
focus on the determination of the DM mass, marginalizing over the rest
of the parameters, i.e., the annihilation branching ratios and the
WIMP-proton cross section.  We leave for future work the study of the
sensitivity of DeepCore to these properties~\cite{extension}.

The main results of this letter are depicted in Fig.~\ref{fig:fig1},
where the relative error in the determination of the DM mass ($m^{\rm
  exp}$ is mass determined by the experiment) is presented for DM
annihilations into $\tau^-\tau^+$ (top panel) and into $b\bar{b}$
(bottom panel).  For each annihilation mode we consider two values for
the WIMP-proton spin-dependent cross section:
$\sigma_{\mathrm{SD}}^\mathrm{p} = 10^{-3}$~pb (in blue), $10^{-4}$~pb
(in orange) for the $\tau^-\tau^+$ channel and
$\sigma_{\mathrm{SD}}^\mathrm{p} = 10^{-2}$~pb (in blue), $4 \times 
10^{-3}$~pb (in orange) for the $b\bar{b}$ channel.  As can be seen
from Fig.~\ref{fig:fig2}, for the $\tau^-\tau^+$ channel and
$m_{\rm{DM}} < 80$~GeV, $\sigma_{\mathrm{SD}}^\mathrm{p} = 10^{-3}$~pb
is excluded at 90\% CL from Super-Kamiokande data~\cite{Kappl:2011kz}.
On the other hand, for the $b\bar{b}$ channel,
$\sigma_{\mathrm{SD}}^\mathrm{p} = 10^{-2}$~pb is also excluded at
90\% CL for some masses in the range depicted from Super-Kamiokande
data~\cite{SKlimit, Kappl:2011kz}.  However, note that these analyses
do not include systematic uncertainties.  In Fig.~\ref{fig:fig1} we
show the results including (light colors) and not including (dark
colors) systematic errors as discussed above.  We can see that if the
WIMP-proton spin-dependent cross section has a value very close to the
current Super-Kamiokande limit, the DM mass could be determined
(including systematic errors) within a $\sim$~50\% uncertainty for
${\rm m_{DM}} < 100$~GeV if DM annihilates dominantly into $b\bar{b}$
or within a few percent for 30~GeV~$\lesssim {\rm m_{DM}} <$~100~GeV
if the dominant DM annihilation channel is $\tau^-\tau^+$.

Systematic uncertainties may have a strong impact on the achievable
precision and their detailed evaluation will play an important role.
In addition to the channels considered here, other channels could be
present, as annihilations directly into neutrino pairs, that would
give rise to a line in the neutrino energy spectrum, leading to the
hardest, and easiest to detect, muon spectra in IceCube/DeepCore and a
better determination of the DM mass.  Conversely, softer channels
would very likely lead to worse results.  Moreover, as the
IceCube/DeepCore neutrino telescope is sensitive to much higher DM
masses, in this case new channels are possible, as annihilations into
gauge bosons or into top quarks.  We leave some of these questions for
future work~\cite{extension}.

{\bf Conclusions.---} 
In this letter we have studied the capabilities of the DeepCore Array
to determine the DM mass in the case of light WIMPs, i.e., 10~GeV~$<
{\rm m_{DM}} <$~100~GeV, by measuring the spectrum of muon-like
events.  We have marginalized over two possible annihilation channels
that we have taken as benchmarks for hard ($\tau^-\tau^+$) and soft
($b\bar{b}$) channels and over the WIMP-proton cross section.  We have
shown that in the case of a cross section close to the current
Super-Kamiokande limits, an excellent measurement of the mass could be
possible (see Fig.~\ref{fig:fig1}).  Therefore the DeepCore Array
provides a new avenue for the determination of the DM mass which is
complementary to other searches~\cite{DMmassgammas, DMmassdirect,
  DMmassneutrinos, Cirelli:2005gh, DMmasscollider} and should be
consistently combined.  This would allow to reduce different sources
of uncertainty and to constrain several DM properties, critical steps
to determine the DM identity.

{\bf Acknowledgments.---}
It is a pleasure to thank D.~Cowen, J.~Koskinen and specially
T.~DeYoung for providing us with useful information about DeepCore,
J.~Edsj\"o for discussions about \texttt{WimpSim}, and T.~A. Girard and
M.~W. Winkler for providing us with the data from SIMPLE and from 
Ref.~\cite{Kappl:2011kz}, respectively.  We also thank C. Orme who
took part in the initial stages of this work.  CRD acknowledges a
scholarship from the Portuguese FCT (ref.~SFRH/BPD/41091/2007).
CRD and SPR are partially supported by the Portuguese FCT through
CERN/FP/116328/2010 and CFTP-FCT UNIT 777, which are partially funded
through POCTI (FEDER).  SPR is also partially supported by the Spanish
Grant FPA2008-02878 of the MICINN.


\vspace{-4mm}

\begin{thebibliography}{srt}

\bibitem{review}
  G.~Jungman, M.~Kamionkowski and K.~Griest,
  Phys.\ Rept.\  {\bf 267}, 195 (1996)
  [arXiv:hep-ph/9506380];
%
  G.~Bertone, D.~Hooper and J.~Silk,
  Phys.\ Rept.\  {\bf 405}, 279 (2005)
  [arXiv:hep-ph/0404175].


\bibitem{Cirelli:2005gh}
  M.~Cirelli {\it et al.}, 
  Nucl.\ Phys.\  B {\bf 727}, 99 (2005)
  [arXiv:hep-ph/0506298].


\bibitem{DMmassgammas}
  S.~Dodelson, D.~Hooper and P.~D.~Serpico,
  Phys.\ Rev.\  D {\bf 77}, 063512 (2008)
  [arXiv:0711.4621 [astro-ph]];
%
  N.~Bernal, A.~Goudelis, Y.~Mambrini and C.~Munoz,
  JCAP {\bf 0901}, 046 (2009)
  [arXiv:0804.1976 [hep-ph]];
%
  T.~E.~Jeltema and S.~Profumo,
  JCAP {\bf 0811}, 003 (2008)
  [arXiv:0808.2641 [astro-ph]];
%
  S.~Palomares-Ruiz and J.~M.~Siegal-Gaskins,
  JCAP {\bf 1007}, 023 (2010)
  [arXiv:1003.1142 [astro-ph.CO]];
%
  N.~Bernal and S.~Palomares-Ruiz,
  arXiv:1006.0477 [astro-ph.HE];
%
  arXiv:1103.2377 [astro-ph.HE].


\bibitem{DMmassdirect}
  A.~M.~Green,
  JCAP {\bf 0708}, 022 (2007)
  [arXiv:hep-ph/0703217];
%
  M.~Drees and C.~L.~Shan,
  JCAP {\bf 0806}, 012 (2008)
  [arXiv:0803.4477 [hep-ph]];
%
  A.~M.~Green,
  JCAP {\bf 0807}, 005 (2008)
  [arXiv:0805.1704 [hep-ph]];
%
  C.~L.~Shan,
  New J.\ Phys.\  {\bf 11}, 105013 (2009)
  [arXiv:0903.4320 [hep-ph]];
%
  L.~E.~Strigari and R.~Trotta,
  JCAP {\bf 0911}, 019 (2009)
  [arXiv:0906.5361 [astro-ph.HE]];
%
  A.~H.~G.~Peter,
  Phys.\ Rev.\  D {\bf 81}, 087301 (2010)
  [arXiv:0910.4765 [astro-ph.CO]];
%
  Y.~T.~Chou and C.~L.~Shan,
  JCAP {\bf 1008}, 014 (2010)
  [arXiv:1003.5277 [hep-ph]];
%
  J.~Billard, F.~Mayet and D.~Santos,
  Phys.\ Rev.\  D {\bf 83}, 075002 (2011)
  [arXiv:1012.3960 [astro-ph.CO]].


\bibitem{DMmassneutrinos}
  J.~Edsjo and P.~Gondolo,
  Phys.\ Lett.\  B {\bf 357}, 595 (1995)
  [arXiv:hep-ph/9504283];
%
  A.~Esmaili and Y.~Farzan,
  JCAP {\bf 1104}, 007 (2011)
  [arXiv:1011.0500 [hep-ph]].


\bibitem{DMmasscollider}
  E.~A.~Baltz, M.~Battaglia, M.~E.~Peskin and T.~Wizansky,
  Phys.\ Rev.\  D {\bf 74}, 103521 (2006)
  [arXiv:hep-ph/0602187];
%
  N.~Alster and M.~Battaglia,
  arXiv:1104.0523 [hep-ex].


\bibitem{Mena:2007ty}
  O.~Mena, S.~Palomares-Ruiz and S.~Pascoli,
  Phys.\ Lett.\  B {\bf 664}, 92 (2008)
  [arXiv:0706.3909 [hep-ph]].


\bibitem{Agarwalla:2011yy}
  S.~K.~Agarwalla, M.~Blennow, E.~F.~Martinez and O.~Mena,
  arXiv:1105.4077 [hep-ph].


\bibitem{Collaboration:2011ym}
  R.~Abbasi {\it et al.} [IceCube Collaboration],
  arXiv:1109.6096 [astro-ph.IM].


\bibitem{Gould92}  
  A. Gould,
  Astrophys.\ J.\ {\bf 388}, 338 (1992).


\bibitem{limitsdirect}
  E.~Aprile {\it et al.}  [XENON100 Collaboration],
  arXiv:1104.2549 [astro-ph.CO];
%
  Z.~Ahmed {\it et al.}  [The CDMS-II Collaboration],
  Science {\bf 327}, 1619 (2010)
  [arXiv:0912.3592 [astro-ph.CO]];
%
  E.~Armengaud {\it et al.}  [EDELWEISS Collaboration],
  arXiv:1103.4070 [astro-ph.CO].


\bibitem{SDvsSD}
  V.~Barger, W.~Y.~Keung and G.~Shaughnessy,
  Phys.\ Rev.\  D {\bf 78}, 056007 (2008)
  [arXiv:0806.1962 [hep-ph]].
%
  G.~Belanger, E.~Nezri and A.~Pukhov,
  Phys.\ Rev.\  D {\bf 79}, 015008 (2009)
  [arXiv:0810.1362 [hep-ph]];
%
  T.~Cohen, D.~J.~Phalen and A.~Pierce,
  Phys.\ Rev.\  D {\bf 81}, 116001 (2010)
  [arXiv:1001.3408 [hep-ph]];
%
  P.~Agrawal, Z.~Chacko, C.~Kilic and R.~K.~Mishra,
  arXiv:1003.1912 [hep-ph].


\bibitem{Felizardo:2011uw}
  M.~Felizardo {\it et al.},
  arXiv:1106.3014 [astro-ph.CO].


\bibitem{SKlimit}
  S.~Desai {\it et al.}  [Super-Kamiokande Collaboration],
  Phys.\ Rev.\  D {\bf 70}, 083523 (2004)
  [Erratum-ibid.\  D {\bf 70}, 109901 (2004)]
  [arXiv:hep-ex/0404025];
%
  T.~Tanaka {\it et al.}  [Kamiokande Collaboration],
  arXiv:1108.3384 [astro-ph.HE].


\bibitem{Kappl:2011kz}
  R.~Kappl, M.~W.~Winkler,
  Nucl.\ Phys.\  {\bf B850}, 505-521 (2011)
  [arXiv:1104.0679 [hep-ph]].


\bibitem{Blennow:2007tw}
  M.~Blennow, J.~Edsjo and T.~Ohlsson,
  JCAP {\bf 0801}, 021 (2008)
  [arXiv:0709.3898 [hep-ph]].


\bibitem{Barger:2007xf}
  V.~Barger, W.~Y.~Keung, G.~Shaughnessy and A.~Tregre,
  Phys.\ Rev.\  D {\bf 76}, 095008 (2007)
  [arXiv:0708.1325 [hep-ph]].


\bibitem{talk}
  T.~DeYoung and J.~Koskinen, private communication. See eg,
  T.~DeYoung talk at RICAP 2011, Rome (Italy), 25-27 May 2011. 


\bibitem{Barr:2006it}
  G.~D.~Barr, T.~K.~Gaisser, S.~Robbins and T.~Stanev,
  Phys.\ Rev.\  D {\bf 74}, 094009 (2006)
  [arXiv:astro-ph/0611266].


\bibitem{Serpico:2010ae}
  P.~D.~Serpico, G.~Bertone,
  Phys.\ Rev.\  {\bf D82}, 063505 (2010)
  [arXiv:1006.3268 [astro-ph.HE]].


\bibitem{extension}
  C.~R.~Das, O.~Mena, S.~Palomares-Ruiz and S.~Pascoli, work in
  progress.


\end{thebibliography}
\end{document}